\documentclass[11pt]{article}

\usepackage{fullpage}
\newenvironment{descit}[1]{\begin{quote} \textit{#1}}{\end{quote}}

\ifx\undefined\psfig\else \fi

\edef\psfigRestoreAt{\catcode`@=\number\catcode`@\relax}
\catcode`\@=11\relax
\newwrite\@unused
\def\typeout#1{{\let\protect\string\immediate\write\@unused{#1}}}
\def\figurepath{./}

\def\@nnil{\@nil}
\def\@empty{}
\def\@psdonoop#1\@@#2#3{}
\def\@psdo#1:=#2\do#3{\edef\@psdotmp{#2}\ifx\@psdotmp\@empty \else
    \expandafter\@psdoloop#2,\@nil,\@nil\@@#1{#3}\fi}
\def\@psdoloop#1,#2,#3\@@#4#5{\def#4{#1}\ifx #4\@nnil \else
       #5\def#4{#2}\ifx #4\@nnil \else#5\@ipsdoloop #3\@@#4{#5}\fi\fi}
\def\@ipsdoloop#1,#2\@@#3#4{\def#3{#1}\ifx #3\@nnil 
       \let\@nextwhile=\@psdonoop \else
      #4\relax\let\@nextwhile=\@ipsdoloop\fi\@nextwhile#2\@@#3{#4}}
\def\@tpsdo#1:=#2\do#3{\xdef\@psdotmp{#2}\ifx\@psdotmp\@empty \else
    \@tpsdoloop#2\@nil\@nil\@@#1{#3}\fi}
\def\@tpsdoloop#1#2\@@#3#4{\def#3{#1}\ifx #3\@nnil 
       \let\@nextwhile=\@psdonoop \else
      #4\relax\let\@nextwhile=\@tpsdoloop\fi\@nextwhile#2\@@#3{#4}}
\newread\ps@stream
\newif\ifnot@eof       
\newif\if@noisy        
\newif\if@atend        
\newif\if@psfile       
{\catcode`\%=12\global\gdef\epsf@start{
\def\epsf@PS{PS}
\def\epsf@getbb#1{%
\openin\ps@stream=#1
\ifeof\ps@stream\typeout{Error, File #1 not found}\else
   {\not@eoftrue \chardef\other=12
    \def\do##1{\catcode`##1=\other}\dospecials \catcode`\ =10
    \loop
       \if@psfile
	  \read\ps@stream to \epsf@fileline
       \else{
	  \obeyspaces
          \read\ps@stream to \epsf@tmp\global\let\epsf@fileline\epsf@tmp}
       \fi
       \ifeof\ps@stream\not@eoffalse\else
       \if@psfile\else
       \expandafter\epsf@test\epsf@fileline:. \\%
       \fi
          \expandafter\epsf@aux\epsf@fileline:. \\%
       \fi
   \ifnot@eof\repeat
   }\closein\ps@stream\fi}%
\long\def\epsf@test#1#2#3:#4\\{\def\epsf@testit{#1#2}
			\ifx\epsf@testit\epsf@start\else
\typeout{Warning! File does not start with `\epsf@start'.  It may not be a PostScript file.}
			\fi
			\@psfiletrue} 
{\catcode`\%=12\global\let\epsf@percent=
\long\def\epsf@aux#1#2:#3\\{\ifx#1\epsf@percent
   \def\epsf@testit{#2}\ifx\epsf@testit\epsf@bblit
	\@atendfalse
        \epsf@atend #3 . \\%
	\if@atend	
	   \if@verbose{
		\typeout{psfig: found `(atend)'; continuing search}
	   }\fi
        \else
        \epsf@grab #3 . . . \\%
        \not@eoffalse
        \global\no@bbfalse
        \fi
   \fi\fi}%
\def\epsf@grab #1 #2 #3 #4 #5\\{%
   \global\def\epsf@llx{#1}\ifx\epsf@llx\empty
      \epsf@grab #2 #3 #4 #5 .\\\else
   \global\def\epsf@lly{#2}%
   \global\def\epsf@urx{#3}\global\def\epsf@ury{#4}\fi}%
\def\epsf@atendlit{(atend)} 
\def\epsf@atend #1 #2 #3\\{%
   \def\epsf@tmp{#1}\ifx\epsf@tmp\empty
      \epsf@atend #2 #3 .\\\else
   \ifx\epsf@tmp\epsf@atendlit\@atendtrue\fi\fi}

\chardef\letter = 11
\chardef\other = 12

\newif \ifdebug 
\newif\ifc@mpute 
\c@mputetrue 

\let\then = \relax
\def\r@dian{pt }
\let\r@dians = \r@dian
\let\dimensionless@nit = \r@dian
\let\dimensionless@nits = \dimensionless@nit
\def\internal@nit{sp }
\let\internal@nits = \internal@nit
\newif\ifstillc@nverging
\def \Mess@ge #1{\ifdebug \then \message {#1} \fi}

{ 
	\catcode `\@ = \letter
	\gdef \nodimen {\expandafter \n@dimen \the \dimen}
	\gdef \term #1 #2 #3%
	       {\edef \t@ {\the #1}
		\edef \t@@ {\expandafter \n@dimen \the #2\r@dian}%
		\t@rm {\t@} {\t@@} {#3}%
	       }
	\gdef \t@rm #1 #2 #3%
	       {{%
		\count 0 = 0
		\dimen 0 = 1 \dimensionless@nit
		\dimen 2 = #2\relax
		\Mess@ge {Calculating term #1 of \nodimen 2}%
		\loop
		\ifnum	\count 0 < #1
		\then	\advance \count 0 by 1
			\Mess@ge {Iteration \the \count 0 \space}%
			\Multiply \dimen 0 by {\dimen 2}%
			\Mess@ge {After multiplication, term = \nodimen 0}%
			\Divide \dimen 0 by {\count 0}%
			\Mess@ge {After division, term = \nodimen 0}%
		\repeat
		\Mess@ge {Final value for term #1 of 
				\nodimen 2 \space is \nodimen 0}%
		\xdef \Term {#3 = \nodimen 0 \r@dians}%
		\aftergroup \Term
	       }}
	\catcode `\p = \other
	\catcode `\t = \other
	\gdef \n@dimen #1pt{#1} 
}

\def \Divide #1by #2{\divide #1 by #2} 

\def \Multiply #1by #2
       {{
	\count 0 = #1\relax
	\count 2 = #2\relax
	\count 4 = 65536
	\Mess@ge {Before scaling, count 0 = \the \count 0 \space and
			count 2 = \the \count 2}%
	\ifnum	\count 0 > 32767 
	\then	\divide \count 0 by 4
		\divide \count 4 by 4
	\else	\ifnum	\count 0 < -32767
		\then	\divide \count 0 by 4
			\divide \count 4 by 4
		\else
		\fi
	\fi
	\ifnum	\count 2 > 32767 
	\then	\divide \count 2 by 4
		\divide \count 4 by 4
	\else	\ifnum	\count 2 < -32767
		\then	\divide \count 2 by 4
			\divide \count 4 by 4
		\else
		\fi
	\fi
	\multiply \count 0 by \count 2
	\divide \count 0 by \count 4
	\xdef \product {#1 = \the \count 0 \internal@nits}%
	\aftergroup \product
       }}

\def\r@duce{\ifdim\dimen0 > 90\r@dian \then   
		\multiply\dimen0 by -1
		\advance\dimen0 by 180\r@dian
		\r@duce
	    \else \ifdim\dimen0 < -90\r@dian \then  
		\advance\dimen0 by 360\r@dian
		\r@duce
		\fi
	    \fi}

\def\Sine#1%
       {{%
	\dimen 0 = #1 \r@dian
	\r@duce
	\ifdim\dimen0 = -90\r@dian \then
	   \dimen4 = -1\r@dian
	   \c@mputefalse
	\fi
	\ifdim\dimen0 = 90\r@dian \then
	   \dimen4 = 1\r@dian
	   \c@mputefalse
	\fi
	\ifdim\dimen0 = 0\r@dian \then
	   \dimen4 = 0\r@dian
	   \c@mputefalse
	\fi
	\ifc@mpute \then
		\divide\dimen0 by 180
		\dimen0=3.141592654\dimen0
		\dimen 2 = 3.1415926535897963\r@dian 
		\divide\dimen 2 by 2 
		\Mess@ge {Sin: calculating Sin of \nodimen 0}%
		\count 0 = 1 
		\dimen 2 = 1 \r@dian 
		\dimen 4 = 0 \r@dian 
		\loop
			\ifnum	\dimen 2 = 0 
			\then	\stillc@nvergingfalse 
			\else	\stillc@nvergingtrue
			\fi
			\ifstillc@nverging 
			\then	\term {\count 0} {\dimen 0} {\dimen 2}%
				\advance \count 0 by 2
				\count 2 = \count 0
				\divide \count 2 by 2
				\ifodd	\count 2 
				\then	\advance \dimen 4 by \dimen 2
				\else	\advance \dimen 4 by -\dimen 2
				\fi
		\repeat
	\fi		
			\xdef \sine {\nodimen 4}%
       }}

\def\Cosine#1{\ifx\sine\UnDefined\edef\Savesine{\relax}\else
		             \edef\Savesine{\sine}\fi
	{\dimen0=#1\r@dian\multiply\dimen0 by -1
	 \advance\dimen0 by 90\r@dian
	 \Sine{\nodimen 0}
	 \xdef\cosine{\sine}
	 \xdef\sine{\Savesine}}}	      

\def\psdraft{
	\def\@psdraft{0}
}
\def\psfull{
	\def\@psdraft{100}
}

\psfull

\newif\if@draftbox
\def\psnodraftbox{
	\@draftboxfalse
}
\@draftboxtrue

\newif\if@prologfile
\newif\if@postlogfile
\def\pssilent{
	\@noisyfalse
}
\def\psnoisy{
	\@noisytrue
}
\psnoisy
\newif\if@bbllx
\newif\if@bblly
\newif\if@bburx
\newif\if@bbury
\newif\if@height
\newif\if@width
\newif\if@rheight
\newif\if@rwidth
\newif\if@angle
\newif\if@clip
\newif\if@verbose
\newif\if@scale
\def\@p@@sclip#1{\@cliptrue}

\def\@p@@sfile#1{\def\@p@sfile{null}%
	        \openin1=#1
		\ifeof1\closein1%
		       \openin1=\figurepath#1
			\ifeof1\typeout{Error, File #1 not found}
			   \if@bbllx\if@bblly\if@bburx\if@bbury
			      \def\@p@sfile{#1}%
			   \fi\fi\fi\fi
			\else\closein1
			    \edef\@p@sfile{\figurepath#1}%
                        \fi%
		 \else\closein1%
		       \def\@p@sfile{#1}%
		 \fi}
\def\@p@@sfigure#1{\def\@p@sfile{null}%
	        \openin1=#1
		\ifeof1\closein1%
		       \openin1=\figurepath#1
			\ifeof1\typeout{Error, File #1 not found}
			   \if@bbllx\if@bblly\if@bburx\if@bbury
			      \def\@p@sfile{#1}%
			   \fi\fi\fi\fi
			\else\closein1
			    \def\@p@sfile{\figurepath#1}%
                        \fi%
		 \else\closein1%
		       \def\@p@sfile{#1}%
		 \fi}

\def\@p@@sbbllx#1{
		\@bbllxtrue
		\dimen100=#1
		\edef\@p@sbbllx{\number\dimen100}
}
\def\@p@@sbblly#1{
		\@bbllytrue
		\dimen100=#1
		\edef\@p@sbblly{\number\dimen100}
}
\def\@p@@sbburx#1{
		\@bburxtrue
		\dimen100=#1
		\edef\@p@sbburx{\number\dimen100}
}
\def\@p@@sbbury#1{
		\@bburytrue
		\dimen100=#1
		\edef\@p@sbbury{\number\dimen100}
}
\def\@p@@sheight#1{
		\@heighttrue
		\dimen100=#1
   		\edef\@p@sheight{\number\dimen100}
}
\def\@p@@swidth#1{
		\@widthtrue
		\dimen100=#1
		\edef\@p@swidth{\number\dimen100}
}
\def\@p@@srheight#1{
		\@rheighttrue
		\dimen100=#1
		\edef\@p@srheight{\number\dimen100}
}
\def\@p@@srwidth#1{
		\@rwidthtrue
		\dimen100=#1
		\edef\@p@srwidth{\number\dimen100}
}
\def\@p@@sangle#1{
		\@angletrue
		\edef\@p@sangle{#1} 
}
\def\@p@@ssilent#1{ 
		\@verbosefalse
}
\def\@p@@sscale#1{
		\def\@p@scale{#1}
		\@scaletrue
}
\def\@p@@sprolog#1{\@prologfiletrue\def\@prologfileval{#1}}
\def\@p@@spostlog#1{\@postlogfiletrue\def\@postlogfileval{#1}}
\def\@cs@name#1{\csname #1\endcsname}
\def\@setparms#1=#2,{\@cs@name{@p@@s#1}{#2}}
\def\ps@init@parms{
		\@bbllxfalse \@bbllyfalse
		\@bburxfalse \@bburyfalse
		\@heightfalse \@widthfalse
		\@rheightfalse \@rwidthfalse
		\@scalefalse
		\def\@p@sbbllx{}\def\@p@sbblly{}
		\def\@p@sbburx{}\def\@p@sbbury{}
		\def\@p@sheight{}\def\@p@swidth{}
		\def\@p@srheight{}\def\@p@srwidth{}
		\def\@p@sangle{0}
		\def\@p@sfile{}
		\def\@p@scost{10}
		\def\@sc{}
		\@prologfilefalse
		\@postlogfilefalse
		\@clipfalse
		\if@noisy
			\@verbosetrue
		\else
			\@verbosefalse
		\fi
}
\def\parse@ps@parms#1{
	 	\@psdo\@psfiga:=#1\do
		   {\expandafter\@setparms\@psfiga,}}
\newif\ifno@bb
\def\bb@missing{
	\if@verbose{
		\typeout{psfig: searching \@p@sfile \space  for bounding box}
	}\fi
	\no@bbtrue
	\epsf@getbb{\@p@sfile}
        \ifno@bb \else \bb@cull\epsf@llx\epsf@lly\epsf@urx\epsf@ury\fi
}	
\def\bb@cull#1#2#3#4{
	\dimen100=#1 bp\edef\@p@sbbllx{\number\dimen100}
	\dimen100=#2 bp\edef\@p@sbblly{\number\dimen100}
	\dimen100=#3 bp\edef\@p@sbburx{\number\dimen100}
	\dimen100=#4 bp\edef\@p@sbbury{\number\dimen100}
	\no@bbfalse
}

\newdimen\p@intvaluex
\newdimen\p@intvaluey
\newdimen\@ffsetvalue
\newdimen\x@ffsetvalue
\newdimen\y@ffsetvalue

\def\compute@offset#1#2{{\dimen0=#1 sp\dimen1=#2 sp
			\advance\dimen1 by -\dimen0
			\dimen1=\sine\dimen1
			\dimen0=\cosine\dimen1
			\ifdim\dimen0<0sp \dimen1=0sp \fi
			\global\@ffsetvalue=\dimen1}}

\def\rotate@#1#2{{\dimen0=#1 sp\dimen1=#2 sp
		  \global\p@intvaluex=\cosine\dimen0
		  \dimen3=\sine\dimen1
		  \global\advance\p@intvaluex by -\dimen3
		  \global\p@intvaluey=\sine\dimen0
		  \dimen3=\cosine\dimen1
		  \global\advance\p@intvaluey by \dimen3
		  }}
\def\compute@bb{
		\no@bbfalse
		\if@bbllx \else \no@bbtrue \fi
		\if@bblly \else \no@bbtrue \fi
		\if@bburx \else \no@bbtrue \fi
		\if@bbury \else \no@bbtrue \fi
		\ifno@bb \bb@missing \fi
		\ifno@bb \typeout{FATAL ERROR: no bb supplied or found}
			\no-bb-error
		\fi
		\if@angle 
			\Sine{\@p@sangle}\Cosine{\@p@sangle}
			\compute@offset{\@p@sbblly}{\@p@sbbury}
			\x@ffsetvalue=\@ffsetvalue
			\compute@offset{\@p@sbburx}{\@p@sbbllx}
			\y@ffsetvalue=\@ffsetvalue

			\rotate@{\@p@sbbllx}{\@p@sbblly}
			\advance\p@intvaluex by -\x@ffsetvalue
			\advance\p@intvaluey by -\y@ffsetvalue
			\edef\@p@sbbllx{\number\p@intvaluex}
			\edef\@p@sbblly{\number\p@intvaluey}

			\rotate@{\@p@sbburx}{\@p@sbbury}
			\advance\p@intvaluex by \x@ffsetvalue
			\advance\p@intvaluey by \y@ffsetvalue
			\edef\@p@sbburx{\number\p@intvaluex}
			\edef\@p@sbbury{\number\p@intvaluey}
			{
			 \count0=\@p@sbbllx \count1=\@p@sbblly
		 	 \count2=\@p@sbburx \count3=\@p@sbbury
			 \dimen0=\@p@sbbllx sp\dimen1=\@p@sbblly sp
		 	 \dimen2=\@p@sbburx sp\dimen3=\@p@sbbury sp
			 \dimen203=\dimen2 \advance\dimen203 by -\dimen0
			 \dimen204=\dimen3 \advance\dimen204 by -\dimen1
			 \ifdim\dimen203<0sp 
			      \count203=\count2 \count2=\count0 
			      \count0=\count203 
			      \global\edef\@p@sbbllx{\number\count0}
			      \global\edef\@p@sbburx{\number\count2}
			 \fi
			 \ifdim\dimen204<0sp 
			       \count204=\count3
			       \count3=\count1
			       \count1=\count204
			       \global\edef\@p@sbblly{\number\count1}
			       \global\edef\@p@sbbury{\number\count3}
			 \fi
			}
		\fi
		\count203=\@p@sbburx
		\count204=\@p@sbbury
		\advance\count203 by -\@p@sbbllx
		\advance\count204 by -\@p@sbblly
		\edef\@bbw{\number\count203}
		\edef\@bbh{\number\count204}
}
\def\in@hundreds#1#2#3{\count240=#2 \count241=#3
		     \count100=\count240	
		     \divide\count100 by \count241
		     \count101=\count100
		     \multiply\count101 by \count241
		     \advance\count240 by -\count101
		     \multiply\count240 by 10
		     \count101=\count240	
		     \divide\count101 by \count241
		     \count102=\count101
		     \multiply\count102 by \count241
		     \advance\count240 by -\count102
		     \multiply\count240 by 10
		     \count102=\count240	
		     \divide\count102 by \count241
		     \count200=#1\count205=0
		     \count201=\count200
			\multiply\count201 by \count100
		 	\advance\count205 by \count201
		     \count201=\count200
			\divide\count201 by 10
			\multiply\count201 by \count101
			\advance\count205 by \count201
		     \count201=\count200
			\divide\count201 by 100
			\multiply\count201 by \count102
			\advance\count205 by \count201
		     \edef\@result{\number\count205}
}
\def\@ScaleInHundreds#1{
		\in@hundreds{#1}{\@p@scale}{100}
		\edef#1{\@result}
}
\def\compute@wfromh{
		\in@hundreds{\@p@sheight}{\@bbw}{\@bbh}
		\edef\@p@swidth{\@result}
}
\def\compute@hfromw{
		\in@hundreds{\@p@swidth}{\@bbh}{\@bbw}
		\edef\@p@sheight{\@result}
}
\def\compute@handw{
		\if@height 
			\if@width
			\else
				\compute@wfromh
			\fi
		\else 
			\if@width
				\compute@hfromw
			\else
				\edef\@p@sheight{\@bbh}
				\edef\@p@swidth{\@bbw}
			\fi
		\fi
}
\def\compute@resv{
		\if@rheight \else \edef\@p@srheight{\@p@sheight} \fi
		\if@rwidth \else \edef\@p@srwidth{\@p@swidth} \fi
}
\def\compute@sizes{
	\compute@bb
	\compute@handw
	\compute@resv
}
\def\psfig#1{\vbox {
	\ps@init@parms
	\parse@ps@parms{#1}
	\compute@sizes
	\if@scale
                \if@verbose
                        \typeout{psfig: scaling by \@p@scale}
                \fi
                \@ScaleInHundreds{\@p@swidth}
                \@ScaleInHundreds{\@p@sheight}
                \@ScaleInHundreds{\@p@srwidth}
                \@ScaleInHundreds{\@p@srheight}
        \fi
	\ifnum\@p@scost<\@psdraft{
		\if@verbose{
			\typeout{psfig: including \@p@sfile \space }
		}\fi
		\special{ps::[begin] 	\@p@swidth \space \@p@sheight \space
				\@p@sbbllx \space \@p@sbblly \space
				\@p@sbburx \space \@p@sbbury \space
				startTexFig \space }
		\if@angle
			\special {ps:: \@p@sangle \space rotate \space} 
		\fi
		\if@clip{
			\if@verbose{
				\typeout{(clip)}
			}\fi
			\special{ps:: doclip \space }
		}\fi
		\if@prologfile
		    \special{ps: plotfile \@prologfileval \space } \fi
		\special{ps: plotfile \@p@sfile \space }
		\if@postlogfile
		    \special{ps: plotfile \@postlogfileval \space } \fi
		\special{ps::[end] endTexFig \space }
		\vbox to \@p@srheight true sp{
			\hbox to \@p@srwidth true sp{
				\hss
			}
		\vss
		}
	}\else{
		\if@draftbox{		
			\hbox{\fbox{\vbox to \@p@srheight true sp{
			\vss
			\hbox to \@p@srwidth true sp{ \hss \@p@sfile \hss }
			\vss
			}}}
		}\else{
			\vbox to \@p@srheight true sp{
			\vss
			\hbox to \@p@srwidth true sp{\hss}
			\vss
			}
		}\fi

	}\fi
}}
\def\psglobal{\typeout{psfig: PSGLOBAL is OBSOLETE; use psprint -m instead}}
\psfigRestoreAt

\newif\ifpdf
\ifx\pdfoutput\undefined
  \pdffalse
\else
  \pdfoutput=1
  \pdftrue
\fi

\ifpdf
  \usepackage[pdftex]{graphicx}
  \usepackage[pdftex]{color}
  \DeclareGraphicsExtensions{.pdf,.png,.jpg}
\else
  \usepackage[dvips]{graphicx}
  \usepackage[dvips]{color}
  \DeclareGraphicsExtensions{.eps,.epsi,.ps}
\fi

\usepackage{times}
\thispagestyle{empty}
\pagestyle{empty}

\def\midv{\mathop{\,|\,}}

\long\def\cbk#1{{\color{red}[CBK: #1]}}
\newlength\colwidth \setlength\colwidth{3.25in}

\title{When being Weak is Brave:\\
Privacy Issues in Recommender Systems}

\author{Naren Ramakrishnan, Benjamin J. Keller, and Batul J. Mirza\\
Department of Computer Science\\
Virginia Tech, VA 24061\\
Email:\{naren,keller,bmirza\}@cs.vt.edu\\
\medskip
\\
Ananth Y. Grama\\
Department of Computer Sciences\\
Purdue University, IN 47907\\
Email: ayg@cs.purdue.edu\\
\medskip
\\
George Karypis\\
Department of Computer Science\\
University of Minnesota, MN 55455\\
Email: karypis@cs.umn.edu}

\begin{document}

\maketitle
\thispagestyle{empty}
\pagestyle{empty}

\begin{abstract}
\noindent
We explore the conflict between personalization and privacy that arises
from the existence of weak ties. A weak tie is an unexpected connection that
provides serendipitous recommendations. However, information about weak ties
could be used in conjunction with
other sources of data to uncover identities and reveal other personal
information. In this article, we use a graph-theoretic model to study the
benefit and risk from weak ties.
\end{abstract}
\newpage

\section{Introduction}
Privacy in Internet services is typically thought of in terms of protecting
attributes of users (and can thus be related to solutions in database security).
However, information provided by a recommender system can also allow the 
privacy of some of its users to be compromised, when used in conjunction with 
other information. For example, consider a system
that recommends books by finding correlations between
a user's ratings and those of other participants for the same books (a
nearest neighbor algorithm~\cite{herlocker}). Suppose as a user, person X
rates only books on computer networking, but has an interest in Indian 
classical music. We could
imagine the following dialogue between person X and the recommender system:

\begin{descit}
{\bf person X:} Would I like ``Evolution of Indian Classical Music''?\\
{\bf recommender:} Yes.\\
{\bf person X:} Why? (surprised)\\ 
{\bf recommender:} People who liked the books you have liked also liked this book.
\end{descit}

\noindent
This is an example of {\it serendipity} in recommendation: person X
does not expect
to receive a recommendation for a book on a topic outside of those
that he/she has rated. But, in fact, it is not by luck that the recommendation 
was provided --- it means that at least one person has rated the book 
in addition to the 
books that person X has rated (and possibly others). The explanation conveys
this, but also indicates that the algorithm used is a nearest neighbor 
algorithm. The serendipity, if person X is malicious, is that he/she has found 
a candidate
{\it weak tie.}

In social network theory, a tie is a relationship between people. The strength 
of a tie is measured in terms of the number of shared associations.
A family is an example: a child knows her mother and her father, and her
parents know each other. Weak ties, on the other hand,
form bridges between two groups of people
who would not otherwise interact. 
Most importantly for us, weak ties allow us to deduce identities. For instance,
one may know that a particular computer science
department has a networking professor
who was an interest in Indian classical music. 
This professor thus provides a tie between 
university faculty and the Indian classical music
aficionados. Consequently, upon meeting the {\it only}
Indian networking professor in that department,
you can safely identify this person as the Indian classical music enthusiast.

A similar situation occurs in recommendation where weak ties provide the
opportunity for serendipitous recommendations. In our example, a candidate
weak tie has been found between networking books and books on Indian
classical music. We don't know if this truly is a weak tie, but can test it by
varying what we rate (the process would involve masquerading as different users
and rating different sets of books to probe around the original ratings).
Through this process, if the query fails for most variations on the ratings,
then we can be more confident that we have a weak tie. And, in fact, the
more restrictive the set of ratings that yield the positive recommendation,
the more confidence we have that there is one person who has rated the
books.

Our ability to find such a minimum set of ratings is where the risk lies
--- we can use this rating set to determine what the other person has rated 
using queries, and perhaps fit this information to our knowledge of people
who use the system. In the end, it is conceivable that we could identify 
our hypothetical Indian music enthusiast/networking professor in the 
recommendation system, and determine what he may have rated. If the system 
allows us to vary the ratings, then we might be able to estimate the 
person's ratings as well.

In this example, a person actually forms a tie between books through their
ratings: a book relates to another if they have both been rated by
someone. The alternate view that we will take is that there is a tie between
two people if they have rated some number of books in common. So, in the
example, the Indian music enthusiast/networking professor has ties to people who
rated networking books and people who rate books on Indian classical music. 
The ties to people who rate Indian classical music books are weak only 
if there are relatively few people who have the same tastes.

Clearly, the task of identifying someone through their ratings is difficult,
and is even harder if more people have similar rating patterns. But the
above example illustrates the risk that may exist even in simple
recommendation. The risk is really to people who would
participate in weak ties, because they are the people who could most easily
be identified. In some application domains (voting preferences and membership
on boards), even knowing that a weak tie 
exists constitutes a breach of privacy.

\subsection*{Our Approach}
Our goal is to model the benefit from and risk to users who participate
in weak ties. In particular, we 
would like to characterize benefits and risks on an algorithm-independent
basis. To achieve this, we use the model of 
`jumping connections' \cite{batul-thesis}
that casts recommendation as making a series of jumps between people, based on 
common ratings (in nearest neighbor recommendation, there is only one jump). 
We describe how this model has relevance to 
conventionally accepted metrics of evaluation (Section~\ref{jc-model}). Using 
this model we can then identify causes for
weak ties in terms of the rating patterns of people (Section~\ref{weak-ties}). 
Furthermore, the model 
allows us to qualify benefits in terms of reachability in a graph, and risk 
in terms of weak ties (Section~\ref{benefits-risks}). Finally 
(Section~\ref{bigger-pictures}), we look at how policies and social structures
can be designed that can support and enable recommender systems.

\section{Recommendation: The Jumping Connections Model}
\label{jc-model}
\begin{figure}
\centering
\begin{tabular}{cc}
& \mbox{\psfig{figure=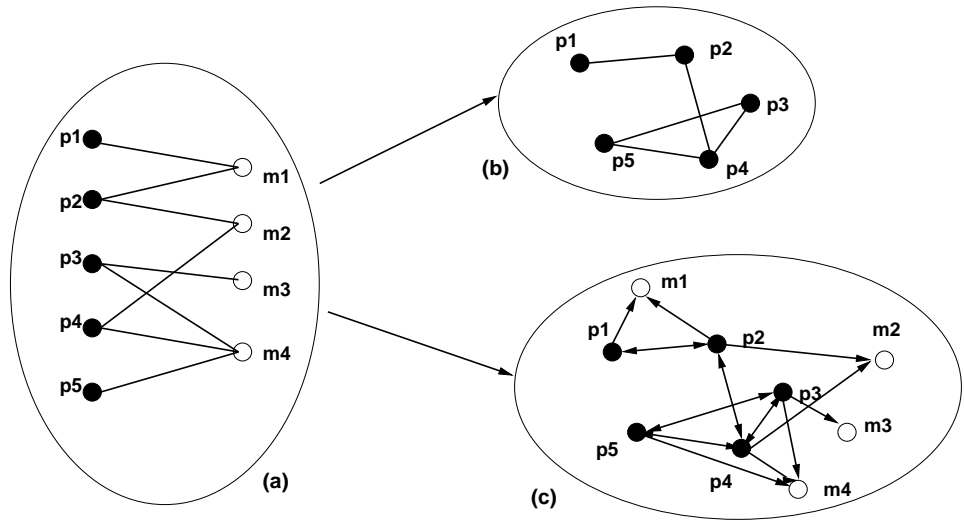,width=5in}}
\end{tabular}
\caption{Illustration of the \emph{skip} jump. (a) bipartite graph
of people and movies. (b) Social network graph, and (c)
recommender graph.}
\label{jc-intro}
\end{figure}

Recommendation algorithms work in a wide variety of ways, from forms of graph
search to learning. This variety presents a difficulty when attempting to
study the risks of recommendation in general. However, we can think of
recommendation as making connections between people who have rated artifacts
in common. We represent people, the artifacts they rated (e.g., movies),
and their ratings as a bipartite graph (Fig.~\ref{jc-intro} (a)). An
algorithm can then {\it jump} over the common artifacts to form a
connection between two people. Fig.~\ref{jc-intro} illustrates a skip
jump where two people are brought together if they rate at least
one movie in common.

A jump induces a {\it social network graph} (Fig.~\ref{jc-intro} (b)),
which includes only people and edges between them (in social network theory,
such a graph shows direct relationships, but here two connected people
need not know one another and their connection depends on the jump). The
{\it recommender graph} (Fig.~\ref{jc-intro} (c)) orients the edges in the social
network graph and adds back the movies. An algorithm can then find paths from
a person making a query to a person who has rated the movie of interest.
Note that Fig.~\ref{jc-intro} illustrates only one way of jumping
--- other jumps are identified by
Mirza \cite{batul-thesis}.

\begin{figure}
\centering
\begin{tabular}{cc}
& \mbox{\psfig{figure=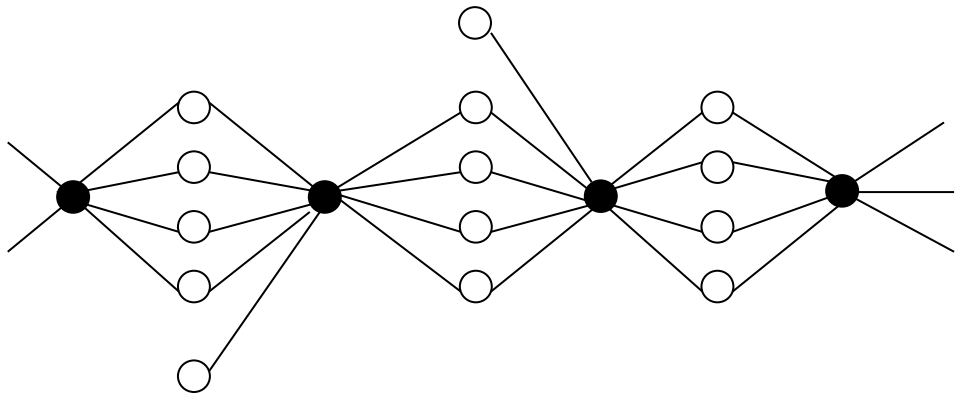,width=5in}}
\end{tabular}
\caption{A path of hammock jumps, with a hammock width
$w=4$.}
\label{hammock-pic}
\end{figure}

In this article, we restrict our attention to {\it hammock} jumps. A hammock
jump of width $w$ connects two people if they have rated at least $w$
movies in common (a skip is a hammock jump of width one). A hammock
path of length $l$ is a sequence of $l$ hammocks, as illustrated in
Fig.~\ref{hammock-pic}. Our hypothesis is that hammock jumps underlie most
recommendation approaches, and at the very least can be used as the basis to
design metrics
for studying privacy issues.
Note that nearest neighbor algorithms (e.g., GroupLens~\cite{konstan1},
LikeMinds, and Firefly) use an implicit hammock sequence of length 1.
The `horting' algorithm of Aggarwal et al.~\cite{aggarwal1} uses 
sequences of explicit hammock-like jumps. 

\begin{figure}
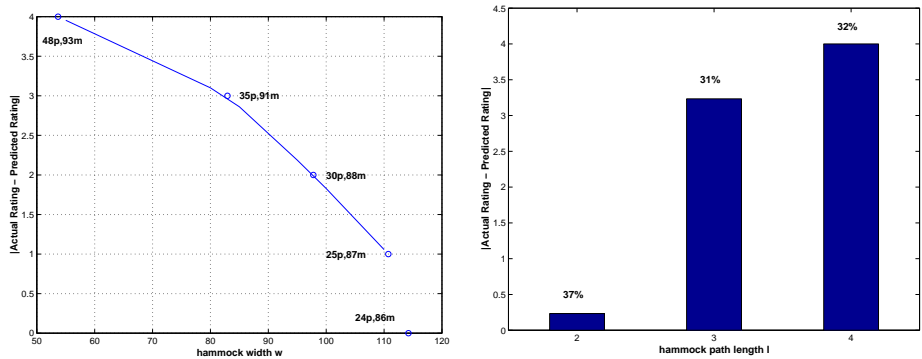

\centering
\begin{tabular}{cc}
\mbox{\psfig{figure=likeminds.epsi,width=2.3in}} & 
\mbox{\psfig{figure=horting.epsi,width=2.3in}}\\
\end{tabular}
\caption{(left) Influence of hammock width $w$ on quality of recommendation.
The annotations denote the fraction of people and movies reachable for
different values of the hammock width. (right) Influence of
hammock path length $l$ on quality of recommendations. The annotations denote
the number of recommendations possible for each value of $l$.}
\label{expt1}
\end{figure}

Our model completely ignores accuracy of predicted values of
ratings, and instead focuses on the parameters of hammock width and
path length. This is because
if recommendation is truly a matter of making (the right) connections, then
a recommendation of a particular movie for a given person can be
characterized by values for the hammock width $w$ and the hammock path 
length $l$. Notice that we do
not emphasize how individual ratings for the nodes (movies) spanning a
hammock are transformed into a prediction.
In addition, it is 
highly likely that there are multiple paths between the same set of
nodes, with various constraints on $w$ and $l$. Intuitively, since
considering more common ratings can be beneficial (see 
\cite{herlocker} for approaches) having a wider hammock could be better (this
has to be carefully done when correlations between ratings are
considered \cite{herlocker}). But if we insist on a wide hammock, we
might have to traverse longer paths to reach a particular
movie from a given person~\cite{batul-thesis}. However, recommendations 
involving shorter
path lengths are preferred, for reasons of explainability, over longer paths.
From a graph-theoretic point of view, $w$ and $l$ thus qualify the
reachability of different movies from a given person, and indirectly provide
a measure of the expected quality of predictions.

Preliminary analysis of the relationship between $w$, $l$, and predictive 
accuracy supports this intuition. Fig.~\ref{expt1} (left) shows a plot
of the average discrepancy between predicted and actual ratings for
each hammock width when using the LikeMinds algorithm (as described 
in~\cite{aggarwal1}). These results were determined by a leave-one-out study,
where an available rating was masked, and a prediction was made for that
rating (using the remaining data). The number of common ratings between
the given person and the person with the highest agreement scalar (and who
contributed to the recommendation) was used as the hammock width. The
results indicate that (for the LikeMinds algorithm), wider hammocks
contribute to better ratings.
Notice that LikeMinds's hammocks do not just model commonality, they also
posit agreement between the rating values spanning a hammock.
While it is certainly true that we can get a poor quality 
recommendation even with a wide hammock (involving perhaps noisy ratings or
a faulty aggregation procedure), 
overall quality of predictions is influenced by greater hammock widths.
However, increasing the hammock width results in a progressive 
disconnection of the social network graph into many components. As a result,
fewer and fewer connections can be made --- Fig.~\ref{expt1} (left) also
lists the fraction of people and movies reachable for various levels of hammock
width. A $w$ of 53 for instance reaches only 48\% of the 
people and 93\% of the movies. By the time the abrupt improvement
in agreement values is observed (after $w \approx 110$), less than 25\%
of the people and only about $86\%$ of the movies are reachable.

Fig.~\ref{expt1} (right) describes the results
of an  experiment where a minimum hammock width constraint was set
at $w=113$ (according to the LikeMinds definition) 
and the resulting recommender graph was analyzed for
paths of varying lengths from people to movies. 
We used the transformation
technique described in \cite{aggarwal1} to make predictions of ratings from
others' ratings, once again using the leave-one-out method. 
Paths in the recommender
graph involve 1, 2, or 3 hops to the person providing a recommendation
and a final hop to the movie being recommended (hence the bucketing of values
into 2, 3, and 4 in Fig.~\ref{expt1}, right).
As can be seen, 
greater lengths (for the same $w$) cause a faster-than-linear
decay in the quality of predictions. We should caution that
horting~\cite{aggarwal1} may exhibit different behavior, though we still
expect longer paths to be of lower quality.

These results support the intuition that wider hammocks and shorter paths
provide better ratings. Hammock widths are determined by rating patterns
that ensure significant overlap. We see this in the 
MovieLens dataset for which each participant rates a minimum of 20 movies
and which has a connected social network graph for all 
$w \le 17$~\cite{batul-thesis}. 

The primary cause of shorter paths is having more connections in the graph.
In the MovieLens dataset, 
a recommendation is almost always possible using
a path of length no longer than 3. This is due to the power-law degree
distribution of the rating patterns. Other graphs, such
as {\it small-world networks} \cite{watts1}, have
small clusters of vertices that are connected by relatively few edges. 
`Weak ties' are important in both these situations
because they make some recommendations possible, and provide others with 
shorter paths. Therefore, weak ties are very important to recommendation.

\section{Ties: Strong, Weak, and Brave}
\label{weak-ties}
In contrast to weak ties\footnote{It is important to note that there is
nothing fundamentally feeble or fragile about a weak tie; a weak tie
creates a powerful and robust link between nodes from different neighborhoods.},
a strong tie connects two people who share 
many associations (like in a family or some other close-knit group).
We can think of weak ties as forming bridges between groups of people
who would otherwise not interact.  Of course, strength and weakness
are relative, and there is no agreed definition of what a strong tie is
in terms of the number of shared associations.

\begin{figure}
\centering
\begin{tabular}{cc}
& \mbox{\psfig{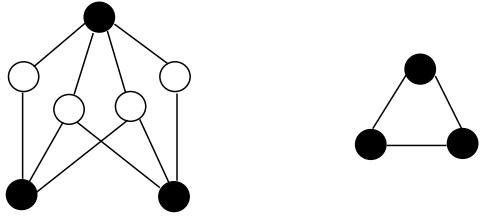}}
\end{tabular}
\caption{Strong ties in a social network graph (right)
induced by a hammock jump on a recommendation dataset (left) with $w=2$.}
\label{figtriad}
\end{figure}

In a graph, strong ties are characterized by a triangle of vertices
(a \emph{triad} in the social network literature).
Fig.~\ref{figtriad}
illustrates how these triads can occur in a social
network graph induced by a hammock jump.
In this case, the width of the hammock jump is 2, and what looks like
two relationships becomes three in the social network graph.
Notice that, in this example, if the hammock jump width were three,
then the resulting social network would not have a triad and so
neither edge would represent a strong tie. 
It is a classical argument in social network theory that
no strong tie can be a bridge
and that two strong ties would imply a third tie~\cite{weak}.

Weak ties are of most interest to us, because they are the foundation
for our notion of risk.
As discussed earlier,
a weak tie in a social setting allows people to identify someone with
other information that they've been given.
Weak ties occur simply because someone knows
someone else outside of their usual circle of friends; or perhaps
there is a person (an `outsider') who is friends with a few people who
each have strong(er) ties to people in different groups.

\begin{figure}
\centering
\begin{tabular}{cc}
\mbox{\psfig{figure=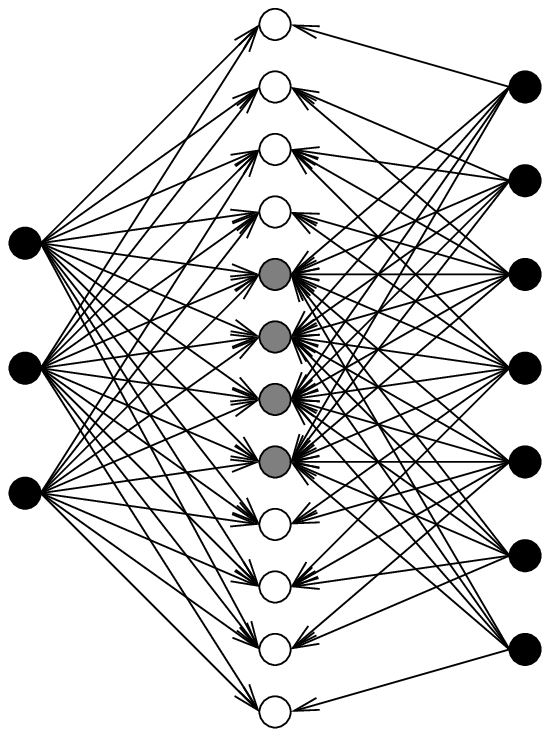,width=2.5in}} &
\mbox{\psfig{figure=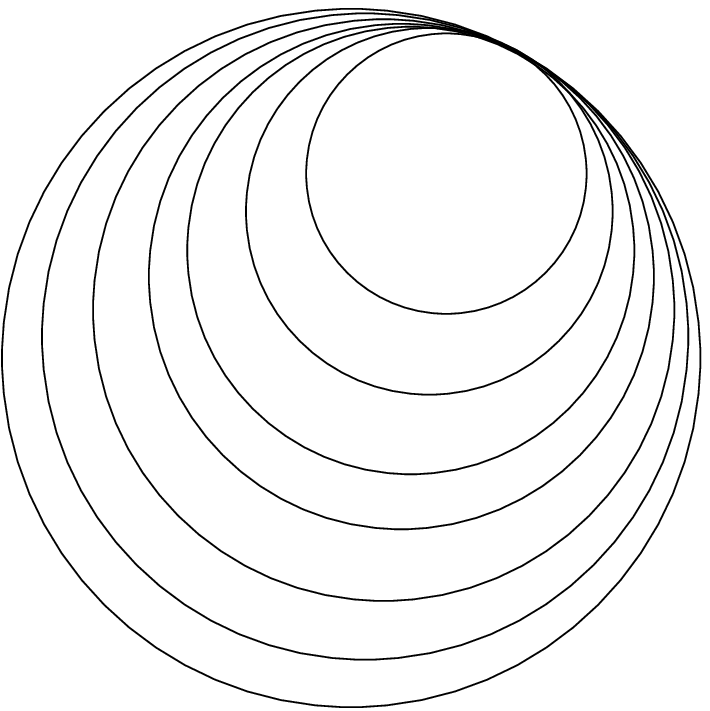,width=2.5in}}
\end{tabular}
\vspace{0.3in}
\begin{tabular}{cc}
\mbox{\psfig{figure=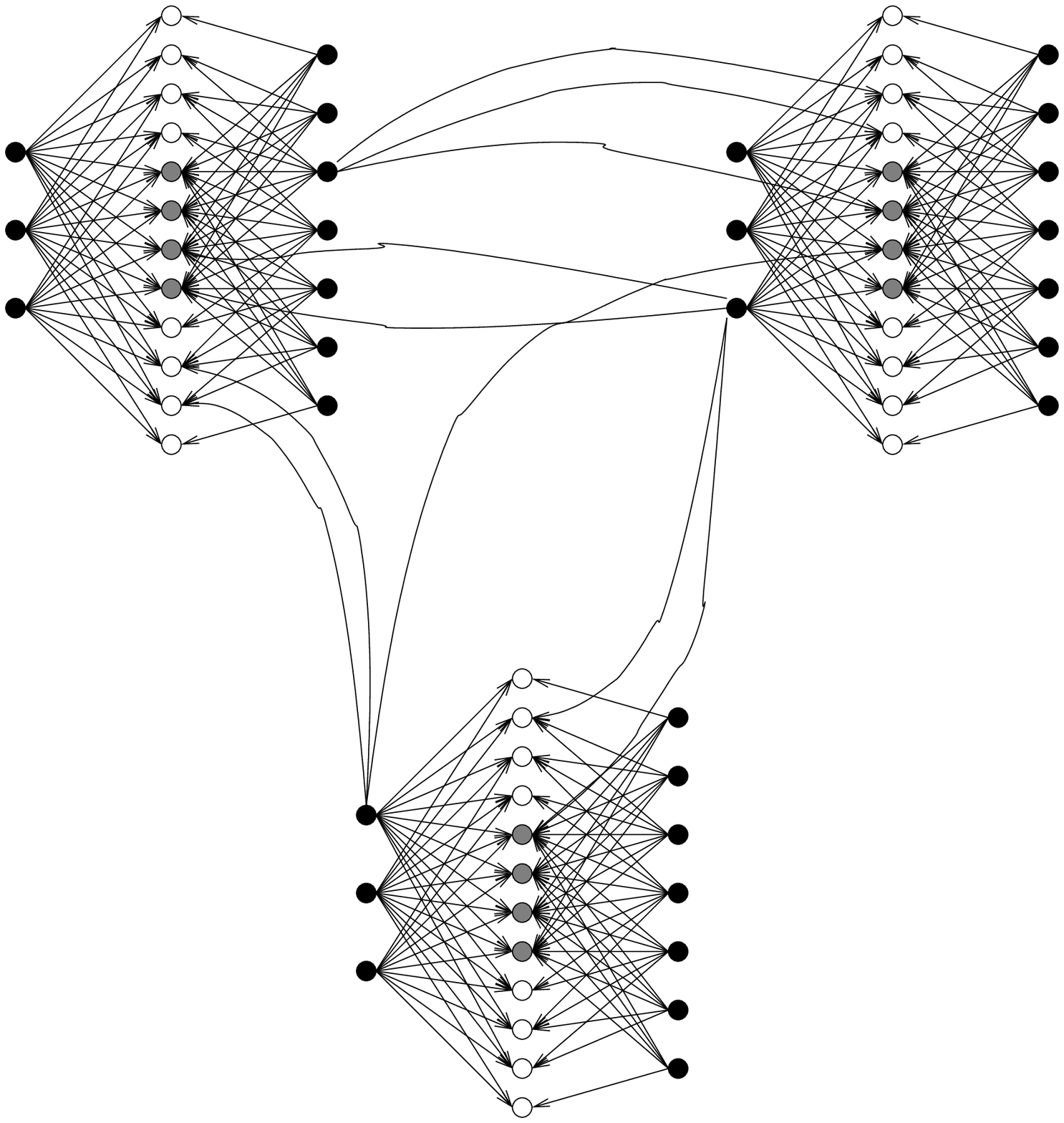,width=2.5in}} &
\mbox{\psfig{figure=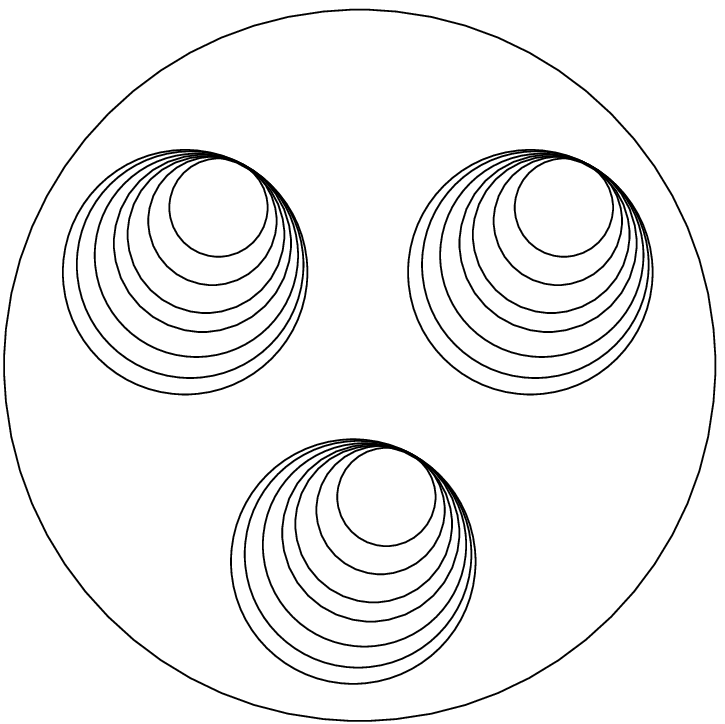,width=2.5in}}
\end{tabular}
\caption{Two different types of induced social networks that can exhibit weak
ties. (top left) A dataset with a power-law induces a low-risk
social network (top right) where increasing hammock widths cause 
a `nested clam shells' picture. Each circle in the social network picture
denotes a group of people brought together. Increasing hammock widths cause
the circles to get progressively smaller.
(bottom left) A dataset with power-laws in only subgraphs and a few weak
ties induces a high-risk social network (bottom right) characterized by
the breakdown of a connected network into disconnected networks.
Some experimental data supporting the
diagrams above can be found in~\cite{batul-thesis}.}
\label{graphs-risks}
\end{figure}

In recommendation, weak ties originate from the rating patterns of the
participants, but the jump process also plays a crucial role. We
hypothesize two fundamental rating patterns. One can be observed
in the public movie recommendation datasets (MovieLens and EachMovie), and
the other is what we would assume for a domain where people have stronger
bias in their tastes (such as books or music).

The movie datasets exhibit a power-law degree distribution as illustrated
in Fig.~\ref{graphs-risks} (top, left). The power-law rating pattern comes from
preferential attachment; for example, some movies (the {\it hits}) are
rated by almost everyone, and some people (the {\it buffs}) rate almost
all movies. Weak ties are rare in this setting but might occur when a person
shows no strong allegiance to any genre and rates relatively few movies in
each (he/she would not be a buff). The real risk to these people is that
they might not have enough ratings in common with anyone so they can be
given recommendations.

The second rating pattern would occur where most people exhibit a preference
for a particular kind of artifact. This is illustrated in Fig.~\ref{graphs-risks} (bottom, left) where there are three subgraphs with power-law structures,
connected by a relatively small number of ratings. This diagram illustrates
one source of weak tie in this setting, which is when someone who
ordinarily only rates artifacts in one domain (e.g., networking books)
rates an artifact in another domain (e.g., Indian classical music books). 
Another
possibility is someone with more eclectic tastes who rates artifacts across
many domains, and unlike in the power-law graph is truly a weak tie. The risk
with weak ties in this rating pattern is that they may allow us to 
identify a person whose ratings can get us from one domain to another.

The jump process, described in Section~\ref{jc-model},
can also create weak ties when using common ratings as the basis for making
connections between people.
Many people might have rated
across several domains, but only a few have enough ratings to satisfy the
jump being used. A final reason relates to merging of data collected from
different settings. For instance, the recent purchase of eToys consumer
data by another retail giant signals the possibility of the creation of
a social network graph with weak ties.

The risk in a weak tie really comes from being the only person with a peculiar
rating pattern --- there is safety in numbers, or at least in
homogeneous tastes (as in power-law graphs).
The more people who rate the same kinds of things, the less likely
that any one of them will be identifiable as participating in a weak tie.
But notice that if the jump definition weeds some of those people out,
the risk is still there (although it is less likely that any additional
information could be used to identify a single person).

\section{The Benefits and Perils of Personalization}
\label{benefits-risks}
Intuitively, a user desires the most benefit from a recommendation that is
based on wider hammocks and shorter path lengths. Of course, to get these 
qualities we have to provide more ratings, and the risk is that we might
introduce a weak tie. The problem then is can we relate how much
we rate to the benefit and risk inherent in recommendation?

\begin{table}
\caption {Movies used in analyzing the benefits of ratings on personalization.
`Star Wars' and `Scream of Stone' had the highest and lowest number of ratings,
respectively.}
\centering
\vspace{0.07in}
\begin{tabular}{|l|r|} \hline\hline
\emph{Movie Name} & \emph {Number of Ratings} \\ \hline
Star Wars & 583\\ \hline
Tommorrow Never Dies & 180 \\ \hline
Robin Hood: Men in Tights & 56 \\ \hline
Scream of Stone & 1 \\ \hline
\end{tabular}
\label{movies-listing}
\end{table}

When there are multiple recommendation paths between a given combination
of person and movie, we would like a benefit formula that captures our
preference for wider hammocks and shorter path lengths. 
By defining the benefit of a recommendation as:
$$\mathrm{benefit} = {w \over{l^2}}$$
we can give
more weight to improvements in path length from $2$ to $1$ than,
say, from $3$ to $2$. This non-linear dependence of quality of interaction
on the length is supported by research in diffusion processes~\cite{watts1}, 
social networks~\cite{weak} and also our own experiments (see Fig.~\ref{expt1},
right). 

We can explore benefit in terms of the number of artifacts that are rated.
Typically, recommender systems require that users rate a minimum number of
artifacts before they can make queries, and so we look at the incremental
benefit received by providing additional ratings. For this purpose, we
analyze the MovieLens dataset where it is required that a user rate 20 movies,
and add a new person. The MovieLens dataset consists of 943 people, 1682
movies, and is connected as a graph.

For the experiment, we introduce a 944th person and incrementally add ratings
from the new person to movies (so that movies with a higher rating
frequency were more likely to be rated). After each rating was added the path
lengths $l$ to particular movies (see Table~\ref{movies-listing}) were computed
for each hammock width $w$. Twenty repetitions were performed for each
additional rating.

\begin{figure}
\centering
\begin{tabular}{cc}
& \mbox{\psfig{figure=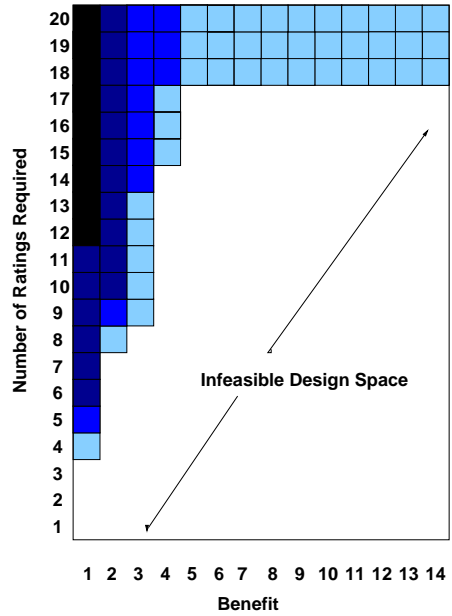,width=2.3in}}
\end{tabular}
\caption{Benefit vs. number of additional ratings required, for various
choices of movie destination nodes. The cells are colored with greater 
intensities corresponding to movies with fewer ratings.}
\label{expt3}
\end{figure}

The benefit from additional ratings for the movies in 
Table~\ref{movies-listing} is shown in Fig.~\ref{expt3}. Each colored cell
indicates that the particular benefit is possible for the corresponding
number of ratings. The feasible benefit regions are actually monotonically
increasing by popularity of the movie --- with more possibilities for
`Star Wars' than for `Tomorrow Never Dies.' The plot shows that if you
want a good recommendation for a less popular movie, you need to provide more
ratings, but can receive good recommendations for popular movies with
fewer ratings. In particular, requesting an improvement in benefit for
a `Star Wars' recommendation from
5 to 14 requires no extra ratings!

\begin{figure}
\centering
\begin{tabular}{cc}
& \mbox{\psfig{figure=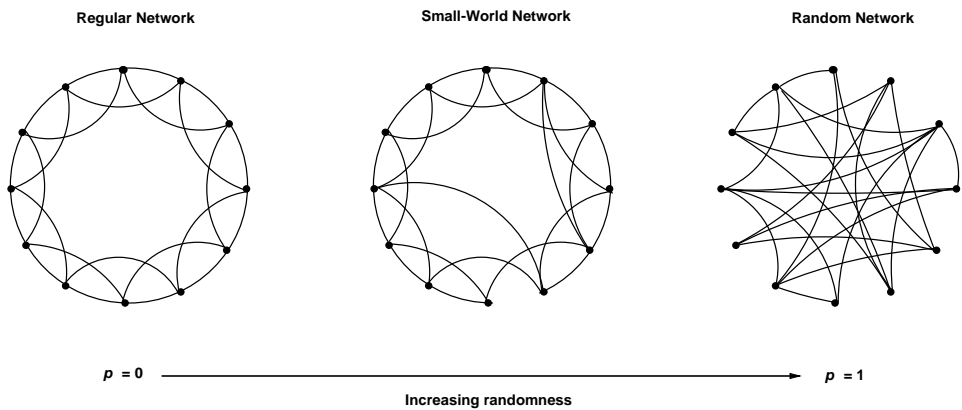,width=5in}}
\end{tabular}
\caption{Random rewiring, starting from a regular wreath network, introduces
weak ties that help model small-world graphs. 
Figure adapted from \cite{watts1}.}
\label{small}
\end{figure}

\begin{figure} \centering
\begin{tabular}{cc} 
\mbox{\psfig{figure=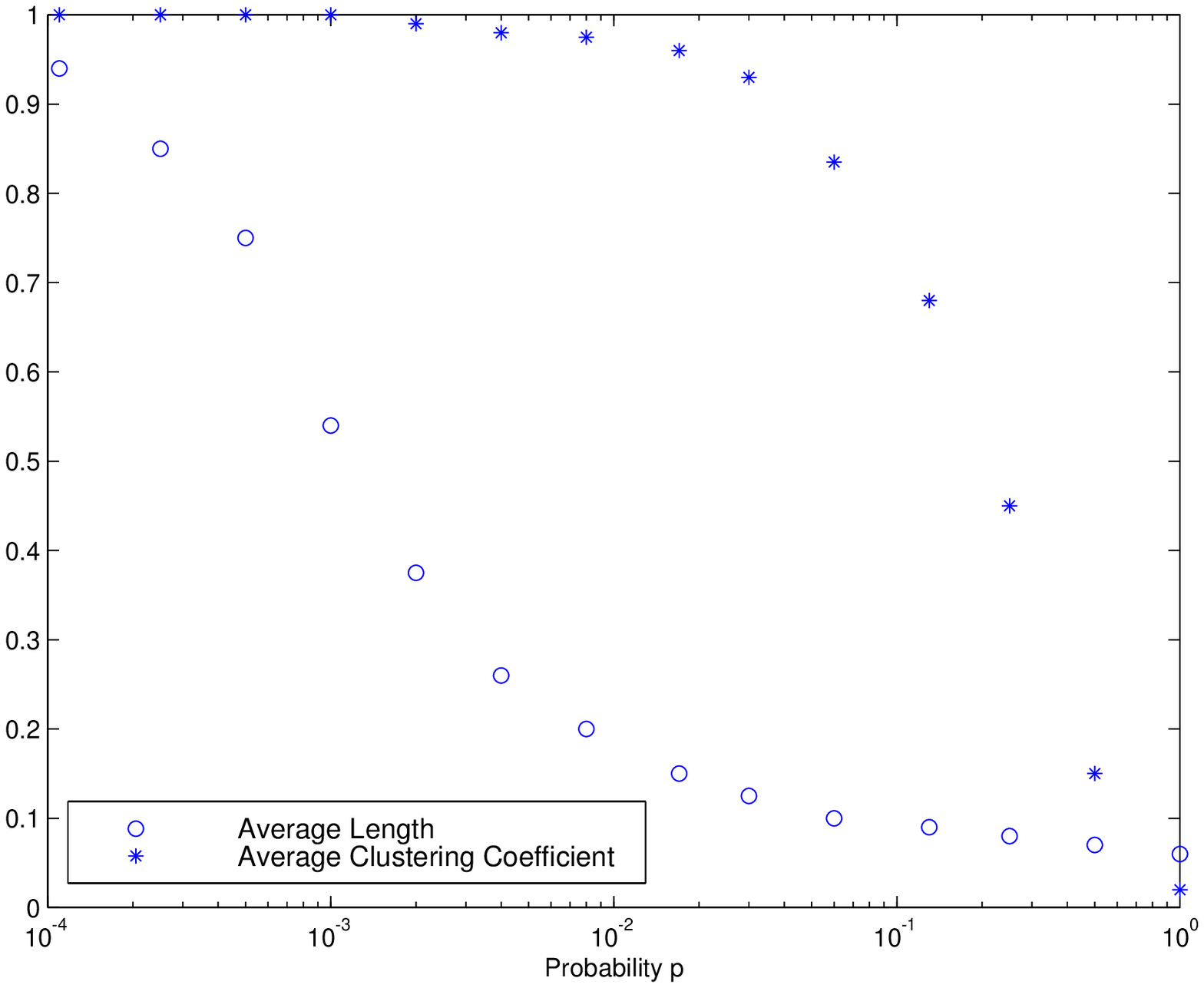,width=2.6in}} &
\mbox{\psfig{figure=expt2.epsi,width=2.6in}}\\
\end{tabular}
\caption{(left) Average path length and clustering coefficient versus
the rewiring probability $p$ (from \cite{watts1}). All measurements are
scaled w.r.t. the values at $p = 0$. (right) Quantifying the risk 
as a function of rewiring probability $p$.} 
\label{smgs}
\end{figure}

The danger involved
in recommendation relates to the probability that a weak connection is
exposed; unfortunately, this is not a static property of a recommendation
path and can only be studied in reference
to the social network graph {\it in the absence} of the considered 
connection. This means that we need a more complete understanding of the
dynamics by which weak ties are introduced, modeled, and employed in a social
network. Such an understanding 
could take the form of a graph-generation
model. Here we use the model of 
Watts and Strogatz \cite{watts1} as a basis for our study of risk.

The intuition is that risk occurs when we have edges that are weak ties between
subgraphs that are cliques (or at least nearly so), and the risk decreases
as more of these edges are added. In particular, the risk is highest when
a new weak tie occurs and the lengths between people decrease dramatically.
As more weak ties are added, the risk decreases.

This idea of risk can be explored in the
Watts-Strogatz model for small-world networks. They show how to
generate a spectrum of graphs from a regular wreath graph by adjusting 
the probability $p$ of rewiring an edge
(Fig.~\ref{small}). When $p$ is zero, we have the wreath; but when $p$ is
one, we have a random graph. The risk from weak ties is low in both the wreath
and random graphs, but increases as the average path length drops but
the vertices are still clustered. Fig.~\ref{smgs} (left) illustrates
the relationship between length and clustering (see~\cite{watts1} for
details of the definitions). When $p$ is between $0$ and $0.1$, the
graph is a small-world network, and poses the most risk from weak ties.

We can express the risk of weak ties in terms of $p$: the risk in having ratings
that form weak ties can be quantified as the rate at which $l$ reduces, as
a function of $p$:
$$\mathrm{risk} = {- {{\partial l} \over {\partial p}}}$$ 

The risk for the dynamics described in Fig.~\ref{small} is given in
Fig.~\ref{smgs}, right (the length values are 
scaled with respect to the length at $p=0$ before calculating the risk). 
Notice that the risk increases 
rapidly (as weak ties
are introduced) and drops down gradually (as more weak ties share
responsibility for length reduction). This captures our intuition pertaining
to disclosure of sensitive information by ferreting out weak ties. 
However, our jumping connections model is not directly parameterized by $p$. 

To be useful, the above formula for risk must relate length reduction to
a metric that could be used to balance personalization and privacy. We
can illustrate the risk of becoming a weak tie by studying what happens
as we decrease the hammock width $w$ in Fig.~\ref{graphs-risks} (bottom). 
Consider the situation when the social network graph is in three disconnected
components. 
Decreasing the hammock width would introduce new edges that
are weak ties, which would contribute to length reduction and thus,
quantification of risk. However, recall that increasing the hammock width is 
desirable from the viewpoint of benefit. Taken another way, benefits improve
monotonically with increasing width $w$ but risk rises rapidly (as 
fewer weak ties share responsibility for length reduction) upto a point and
then drops sharply. 

\begin{figure}
\centering
\begin{tabular}{cc}
& \mbox{\psfig{figure=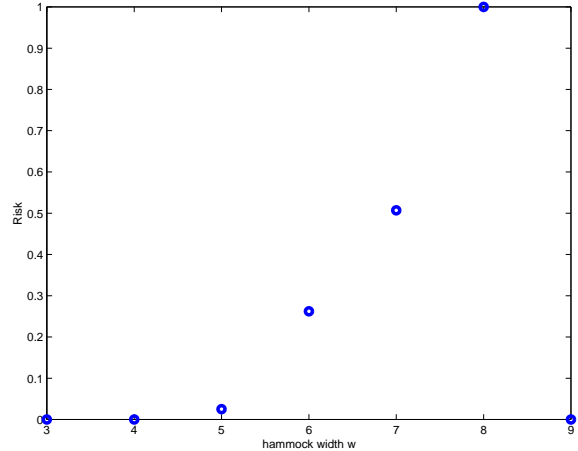,width=3in}}
\end{tabular}
\caption{Risk as a function of hammock width $w$.}
\label{expt4}
\end{figure}

To explore this setting, we created an artificial dataset that
consists of three subgraphs with power-law degree distributions, each
with 200 people and 75 artifact vertices. Each person node is linked to
at most 15 artifact nodes within the same subgraph. Specifically, the
people and artifacts were ordered, and the $b^{th}$ person rated the first
$\lceil 75 b^{-\epsilon} \rceil$ artifacts.
The value of $\epsilon$
was calibrated to achieve a minimum rating of $15$ artifacts. Then three
extra people were added who rate 
(at most 15)
artifacts in all three connected components, again with a 
`master' power-law. 

For a hammock width of 9, the social network of this graph
consists of three disconnected components. By decreasing the hammock
width, weak ties will be introduced into the social network, and the
path lengths decrease. The results are plotted in Fig.~\ref{expt4}
(lengths are scaled against the path length for $w=8$). As could
be expected, risk is highest when the graph is first connected.

It is not possible to provide a traditional benefit-risk profile, as is
customary in analysis. This is because recommender systems aggregate the
ratings of many participants when computing a recommendation. A user's
benefits comes from `plugging into' the social network by 
providing a sufficient number of ratings, but a user's risk depends
not only on what is rated but also
on what other people rate. Ultimately, the difficulty
comes from the fact that risk occurs even if recommendation queries are
not made, but benefit requires that the user make queries. 

The two qualitative conclusions from our studies are that (i) a few 
weak ties are more risky than a lot of weak
ties, and (ii) more so, in some (induced) social networks than others.

\section{Concluding Remarks}
\label{bigger-pictures}
The very factors that make weak ties useful are the ones that 
raise the threat of privacy. We have demonstrated that under certain conditions,
recommendations could involve weak ties and could potentially compromise
the privacy of individuals. Like most problems in computer security, the
ideal deterrents are better awareness of the issues and more openness in 
how recommender systems operate in the market place. In particular, policies
and methodologies employed by an individual site should be made clear. 
Sites that involve multiple homogeneous networks have a crucial responsibility
in clarifying the role of weak ties in their system designs and what forms of
mechanisms are in place to thwart hackers.

Ideally, recommender systems should convey to the user both benefits and risks 
in an intuitive manner. One possibility is to present the user with plots of
benefit and risk versus user-modifiable parameters --- ratings, $w$, and $l$
(if the algorithm allows their direct specification). Another possibility is
to qualify the risks and benefits associated with rating each individual 
artifact (as a function of the previous ratings in the system). Providing
a rating for `Scream of Stone' for instance would provide dramatic improvements
in benefit than providing a rating for `Star Wars.' At the same time, 
the system should qualify the extent to which a user becomes a weak tie,
by such a rating.

Singh and colleagues \cite{singh-cacm} make a provoking observation in
drawing comparisons from community-based networks to recommender systems ---
namely, that people really want to control to whom they reveal their ratings
but would like to know how recommendations are being made. In a distributed
setting, one can imagine a scenario where people specify how data collected
from their interactions should be modeled and used. Interfaces for 
privacy management are woefully inadequate and their role is only now 
being recognized \cite{etzioni-cacm}. Extending the results here to 
a distributed setting where people can set arbitrary constraints on their
station in the social network graph (whether they are willing to participate
in a path?; are there constraints on such participation?; would they provide
ratings if they knew that it would contribute to a weak tie?) is a possible
direction for future research. 

One wonders if weak ties will happen at all, if concerns are raised about
their compromise. Social network theory postulates that they are
the primary mechanisms by which micro-level interactions can manifest at
macro levels, and that such ties will be kindled whenever communities have
to be mobilized for collective action. It remains to be seen if weak ties
induced by jumps in a recommender system also conform to similar 
distributed organization. 
 
\bibliographystyle{plain}
\bibliography{ppp}

\end{document}